\documentclass[aps, twocolumn, pra, showpacs]{revtex4}
\usepackage{graphicx}
\usepackage{amsmath}
\usepackage{amsfonts}
\usepackage{amssymb}
\usepackage{epsfig}

\begin{document}

\title{Super Mario's prison break --- a proposal of object-intelligent-feedback-based \\
classical Zeno and anti-Zeno effects}
\author{Shi-Jian Gu} \email{sjgu@phy.cuhk.edu.hk}
\affiliation{Department of Physics and ITP, The Chinese University of Hong
Kong, Hong Kong, China}

\begin{abstract}
Super Mario is imprisoned by a demon in a finite potential well. He can escape
from the well with the help of a flight of magic stairs floating in the space.
However, the hateful demon may occasionally check his status. At that time, he
has to make a judgement of either jumping to the inside ground immediately in
order to avoid the discovery of his escape intention, or speeding up his escape
process. Therefore, if the demon checks him too frequently such that there is
no probability for him to reach the top of the barrier, he will be always
inside the well, then a classical Zeno effect occurs. On the other hand, if the
time interval between two subsequent checks is large enough such that he has a
higher probability of being beyond the demon's controllable range already, then
the demon's check actually speeds up his escape and a classical anti-Zeno
effect takes place.
\end{abstract}

\date{\today }
\maketitle




The fletcher's paradox proposed by Zeno of Elea states that a flying arrow is
motionless because it occupies an equal space when it is at rest and will be
occupying an equal space in locomotion at any moment. The paradox is clearly
impossible in Newtonian mechanics because our common sense tells us that the
size of the arrow is fixed without any fluctuation and there is no projective
measurement in Newtonian mechanics. Nevertheless, quantum mechanics provides
such a possibility of \textquotedblleft motionlessness". That is an unstable
quantum state would never decay if it is observed continuously. This
phenomenon, called quantum Zeno effect, was first proposed by Misra and
Sudarshan \cite{BMisra77} in 1977. There are two key ingredients in the quantum
Zeno effect. They are the informatic description of quantum states and the
projective measurement. Both of them are among the fundamentals of quantum
mechanics. Because of the absence of the projective measurement in classical
physics, the quantum Zeno effect (see a review \cite{KKoshinoPhysRep}) then
becomes a fascinating game existing uniquely in the framework of quantum
mechanics.

To have a clear picture and for the later comparison, we first give a simple
prove of the quantum Zeno effect. Consider a general quantum system described
by a Hamiltonian $H$, and suppose the system is initially at an excited state
$\left\vert \Psi (0)\right\rangle$, the quantum state, according to quantum
mechanics, will evolve like
\begin{equation}
\left\vert \Psi (t)\right\rangle =e^{-iHt}\left\vert \Psi (0)\right\rangle.
\end{equation}
If we measure the state after a small time interval $\delta t$, the survival
probability of $\left\vert \Psi (0)\right\rangle $ can be measured by
\begin{equation}
P=\left\vert \langle \Psi (0)\left\vert \Psi (\delta t)\right\rangle
\right\vert \simeq 1-\frac{(\delta t)^{2}}{2}\bar{H}^{2},
\end{equation}%
where $\bar{H}^{2}$ denotes the fluctuation of $H$. Therefore, if we perform
$N$ measurements during a fixed time interval $\Delta t=N\delta t$, the
final survival probability of $\left\vert \Psi (0)\right\rangle $ becomes
\begin{equation}
P\simeq 1-\frac{\left( \Delta t\right) ^{2}\bar{H}^{2}}{2N}
\end{equation}%
which tends to 1 if $N\longrightarrow \infty $. This concludes a simple prove
of the quantum Zeno effect, i.e. the state $|\Psi (0)\rangle$ will never decay
if it is observed continuously. The quantum anti-Zeno effect
\cite{Kofman96,Kofman00} takes place if $\delta t$ is larger enough, then
$|\Psi (\delta t)\rangle $ is far away from $|\Psi (0)\rangle $, for instance
$P<1/2$, then any projective measurement will speed up the decay of the state
$|\Psi (0)\rangle $ to other states.

\begin{figure}[tbp]
\includegraphics[width=8cm]{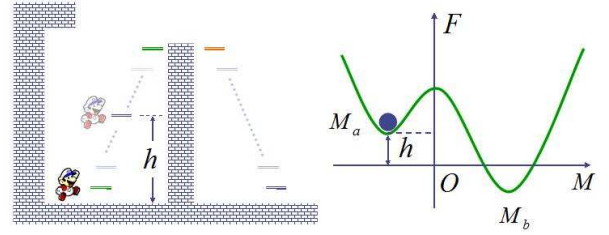}
\caption{LEFT: Super Mario is imprisoned in a finite potential well and he want
to escape from the well. RIGHT: An effective model for Super Mario's escape ---
a transition from an initial metastable state at $M_a$ to a final stable state
at $M_b$ with a certain probability.} \label{fig:mario.eps}
\end{figure}

The quantum Zeno and anti-Zeno effects show how environmental interferences
speed up or slow down a physical process. One may wonder why such a paradigm
becomes a paradox in classical physics. Actually, the first ingredient in the
above Zeno effect is not unique in quantum mechanics. The fluctuation term can
be realized via a statistical process though in this case it might be classical
one while the quantum Zeno effect is based on quantum fluctuation. However, the
second ingredient about the projective measurement does not exist in a
classical physics process. So it is almost impossible to image Zeno and
anti-Zeno effects in classical physics. However, as we will show below, a
projective measurement in classical world can be also realized by an inclined
judgement made by the measured object. That is an environmental interference
can lead to an intelligent feedback from the measured object. Though such a
feedback is somehow beyond physical law, it provides a possibility of Zeno and
anti-Zeno effects in our classical world.

We consider such a scenario about Super Mario's \cite{Mario} prison escape
\cite{prisonescape}. He is imprisoned by a demon in a finite potential well,
and a flight of magic stairs, which can provide him a random potential to jump
up or down with a certain probability distribution, floats in the space. Super
Mario can escape from the well with his own efforts and the random potential
provided by the stairs. His effort is to block up his standing height with
gradual progress hence enhances his chance continuously to reach the top of
barrier. The random potential ensures that his state is a probability
distribution within a certain time interval. Therefore, he can finally escape
from the well given the duration time is longer enough. However, the hateful
demon feels worried about him and may occasionally check his status. Once Super
Mario hears the demon's step, he has to make a judgement of either jumping to
the inside ground immediately in order to avoid the discovery of his escape
intention, or speeding up the escape process. His judgement is made based on
the effective distance of his current state from the initial state because the
distance determines his chance of final successful escape. Therefore, if the
demon checks him too frequently such that there is no probability for him to
even reach the top of the barrier, he will be always inside the well, then a
classical Zeno effect occurs. On the other hand, if the time interval between
two subsequent checks is large enough such that he might have already been on
those steps outside the well, then the demon's check actually speeds up his
escape and a classical anti-Zeno effect takes place. Such a scenario, though is
beyond physical laws in physical textbooks, seems reasonable in our daily life.

Let us now simply compare the above classical Zeno effect with the quantum one.
First, in both effects, the fluctuation plays an important role. In the
classical Zeno effect, it is a classical fluctuation, while in the quantum Zeno
effect, it becomes a quantum fluctuation. Both fluctuations ensure an
algebraical decay of the physical state at the very beginning time. Meanwhile,
we would like to point out also that the algebraical decay can be realized even
without fluctuations, which therefore, is not a necessary condition. Secondly,
the significant difference between both effects is the definition of the
projective measurement. In quantum mechanics, the measurement performed in
experiments is artificial, while the projection of quantum state is a postulate
whose correctness is assumed to be checked by experiments only. In our
scenario, how to measure to the object's state is artificial (done by the
demon), the ``projective" output is also artificial (an intelligent feedback by
Super Mario). The later is possible if and only if the measured object itself
has an inclined selective response. Therefore, the ``projection" of Super Mario
to hide his escape or speed up his escape is somehow beyond physical science,
but belongs to behavioral science. From this point of view, the classical Zeno
and anti-Zeno effects we proposed here look like a combination of physical
science and behavioral science.

The above game provides us only a sketch for the classical Zeno and anti-Zeno
effects. To quantitatively illustrate how Super Mario escapes from the well, we
need to establish an effective model to describe his process. Actually, Super
Mario's escape process is very similar to the decay of a classical metastable
state to a final stable state in statistical physics except for that the
projective feedback should be done artificially. That is the demon's check and
Super Mario's projective feedback can be simulated by projecting artificially
the system's state to the original one. Then we can use a statistical system,
like the two-dimensional Ising model as an effective model to clarify the
classical Zeno and anti-Zeno effects.

\begin{figure}[tbp]
\includegraphics[width=8cm]{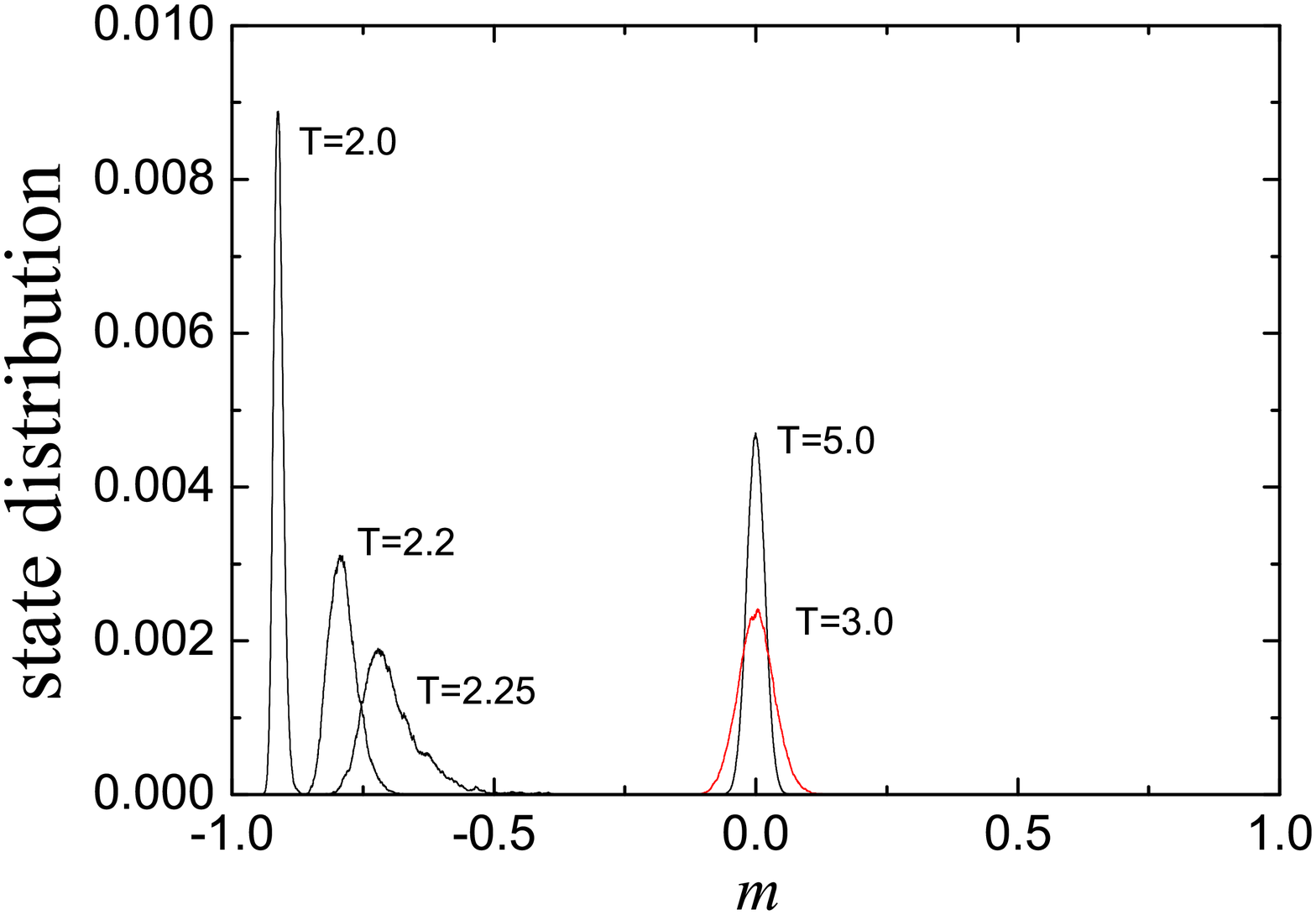}
\caption{Thermal states of the Ising model in the magnetization space for a
sample of $N=100\times 100$ at $T=2.0, 2.2, 2,25, 3, 5$ respectively. Each
state is normalized from $2\times 10^8$ importance samplings via the Metroplis
algorithm.} \label{figure32.eps}
\end{figure}

The Hamiltonian of the two-dimensional Ising model defined on a square
lattice reads
\begin{equation}
H=-\sum_{\left\langle ij\right\rangle }\sigma _{i}^{z}\sigma
_{j}^{z}-h\sum_{i}\sigma _{i}^{z},
\end{equation}
where $\sigma _{i}^{z}=\pm 1$ denotes the spin state at site $i$, $h$ is the
magnetic field, and $M=\sum_i \sigma _{i}^{z}$ is the magnetization. The model
has been solved exactly by Onsager \cite{LOnsager1944} in 1944. According to
the exact solution, a thermal phase
transitions occurs at $T_{c}\simeq 2.269185$ in the thermodynamic limit \cite%
{BMMccoyb}. The system is in a paramagnetic state above the critical point. The
thermal distribution is located in the middle of the magnetization space, and
there is no symmetry breaking. Below the critical point, the system will be
self-magnetized and be in an ordered phase with either positive $M$ or negative
$M$. A sketch of thermal distributions at various temperatures is shown in Fig.
\ref{figure32.eps}, which is simulated from a $100\times 100$ sample and
$2\times 10^8$ importance samplings.

In order to study the classical Zeno effects, we focus on the ordered phase
below the critical point for \emph{a finite system}. To be precise, we will
focus only on a finite $100\times 100$ sample in this paper to ensure a finite
barrier between two magnetized states. In terms of the free energy, there are
two wells separated by a finite barrier in the magnetization space, and Super
Mario is supposed to be in one of the two wells. Since the barrier is finite,
Super Mario has a certain probability, though very very small, to finally
transit to the another well. The probability is determined by both the
Boltzmann distribution and density of state in the magnetization space. We
assume that the magic stair shown in the left picture of Fig.
\ref{fig:mario.eps} offers such a probability to Super Mario to jump up or
down. Moreover, he makes also his own effort, say blocking up his starting
point $h$ (corresponding to magnetic field) with some boxes, to enhance his
potential hence the transition probability to the right well. So if we plot the
free energy in the magnetization space, it looks schematically like the right
picture of Fig. \ref{fig:mario.eps}.

\begin{figure}[tbp]
\includegraphics[width=8cm]{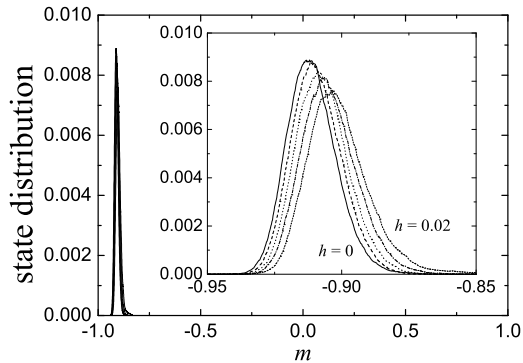}
\caption{Thermal states of the Ising model in the magnetization space for a
finite sample of $N=100\times 100$, $\protect\beta=0.5$, and various $h=0,
0.005, 0.010, 0.015, 0.020$. Each state is normalized from $2\times 10^8$
importance samplings via the Metroplis algorithm.} \label{statefigure.eps}
\end{figure}

\begin{figure}[tbp]
\includegraphics[width=8cm]{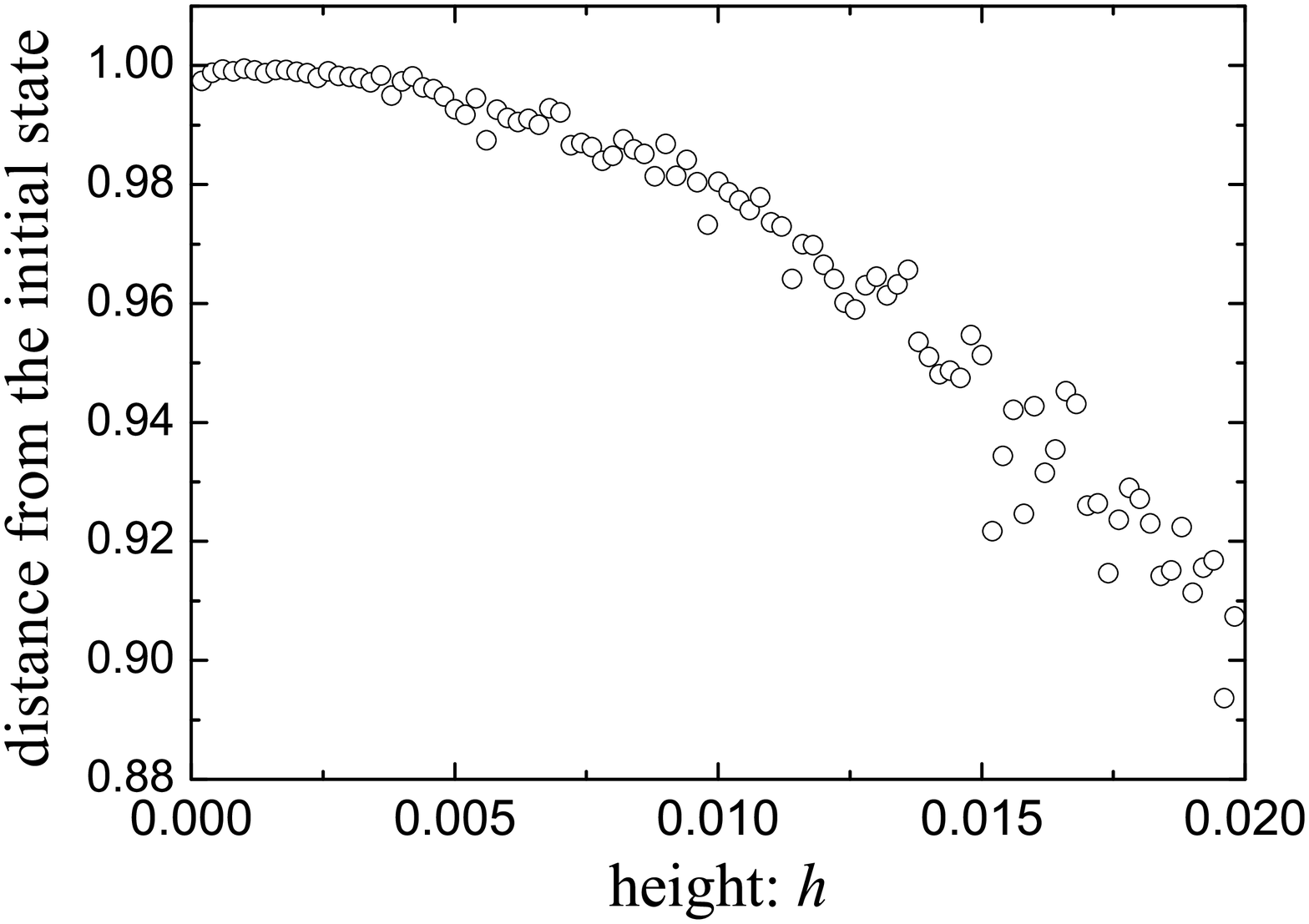}
\caption{The thermal distance, i.e, fidelity, of the thermal state at $h$
from the initial state as a function of $h$.}
\label{fidelityfigure.eps}
\end{figure}

We assume Super Mario blocks up his starting point $h$ so slowly hence the
adiabatic theorem works. He will be in a metastable state as $h$ increases
until a complete transition occurs. In Fig. \ref{statefigure.eps}, we show the
state of Super Mario with various small $h$s. From the figure, we can see the
state of Super Mario changes from his initial position to the right well
gradual. In Fig. \ref{fidelityfigure.eps}, we show his distance from the
initial state as a function of $h$. Therefore, it is certainly expected that
Super Mario will finally escape to the right well if there is no other
interference. However, if the demon checks his status after a time interval
$\delta t$, then Super Mario should make a decision to return to the initial
state or speed up his escape. Her decision depends on the distance between the
current state and the initial state. The distance can be measured by the
fidelity between two thermal states \cite{PzanardiTF}.

To be precise, according to statistical physics, Super Mario's initial state
(with symmetry breaking) can be described by
\begin{equation}
\rho (T,h=0)=\frac{1}{Z}\sum_{M<0}e^{-\beta E_{M}}
\end{equation}%
where the partition function is
\begin{equation}
Z=\sum_{M<0}e^{-\beta E_{M}}.
\end{equation}%
When he blocks up his height with a velocity of $v$, which corresponds to the
increasing rate of the external magnetic field, i.e. $h=vt$, his instant state
becomes
\begin{equation}
\rho (T,h)=\frac{1}{Z}\sum_{M<0}e^{-\beta (E_{M}-hM)}.
\end{equation}%
The state has clearly been dragged from the initial state under the external
field, as shown in Fig. \ref{statefigure.eps}. In the parameter space of $h$%
, the distance between the initial state and the instant state at $h$ can be
measured by the thermal-state fidelity \cite{PzanardiTF},
\begin{equation}
F(0;h)=\frac{1}{\sqrt{Z(0)Z(h)}}\sum_{M<0}e^{-\beta (2E_{M}-hM)/2}.
\end{equation}%
Then after a time interval $\delta t$,
\begin{equation}
F(0; \delta h)\simeq 1-\frac{(\delta h)^{2}}{2}\chi _{F}.
\label{eq:fidelitydef}
\end{equation}%
where $\chi _{F}$ is the thermal-state fidelity susceptibility which is
defined as \cite{WLYou}
\begin{equation}
\chi _{F}=\frac{\beta ^{2}\left( \left\langle M^{2}\right\rangle
-\left\langle M\right\rangle ^{2}\right) }{4}.
\end{equation}%
The fidelity susceptibility is proposed in recent studies on the fidelity
approach to quantum phase transitions \cite{Zanardi06} (See also a review
article \cite{SJGuReview}). In quantum mechanics, the fidelity susceptibility
denotes the leading response of a quantum state to an adiabatic parameter
\cite{WLYou}. The most important here is that the leading term in fidelity is
quadratic. In Fig. \ref{fidelityfigure.eps}, we
show the fidelity defined in Eq. (\ref{eq:fidelitydef}) as a function of $%
\delta h$. Clearly, though there are some fluctuations, the fidelity is almost
parabolic, as being consistent with Eq. (\ref{eq:fidelitydef}).

Now we assume that the demon checks Super Mario's status $N$ times uniformly in
a time interval $\Delta t=t_{f}-t_{i}$. That is the time interval between two
subsequent checks is
\begin{equation}
\delta t=\frac{\Delta t}{N},
\end{equation}%
then
\begin{equation*}
F(0; \delta h)=1-\frac{v^{2}\left( \Delta t\right) ^{2}\chi _{F}}{2N^{2}},
\end{equation*}
which defines also the probability that Super Mario want to hide himself. Then
the final survival probability, after the time interval $\Delta t$
, becomes%
\begin{equation*}
P=\left[ 1-\frac{\left( v\Delta t\right) ^{2}\chi _{F}}{2N^{2}}%
\right] ^{N}.
\end{equation*}%
In the continuous limit $N\rightarrow \infty $,
\begin{equation*}
P \simeq 1 -\frac{\left( v\Delta t\right) ^{2}\chi _{F}}{2N } ,
\end{equation*}%
which becomes 1, then a classical Zeno effect takes place.

On the other hand, if the duration time $\delta t$ between two subsequent
checks is very large, say $F(0,\delta h)>1/2$, then Super Mario might judge
that he has a very high probability to escape from the well successfully. He
will speed up his escape process when he hears the demon's step. A classical
anti-Zeno effect then takes place.

In summary, we have presented a gedanken proposal of classical Zeno and
anti-Zeno effects via a scenario of Super Mario's escape progress. Bearing some
analogy to the quantum Zeno and anti-Zeno effects, environmental interferences
can either slow down or speed up such a classical progress. The significant
difference between the quantum and classical Zeno effects is how to understand
the projective feedback from the measured object. In the quantum Zeno effect,
the projective measurement is a postulate; while in our classical Zeno effect,
it is somehow an inclined feedback which is intelligent. Therefore, we can call
these newly proposed effects as object-intelligent-feedback-based classical
Zeno and anti-Zeno effects.

Finally, we would like to emphasize that the effects we proposed here is very
popular in our daily life. An apparent example is: in order to prevent some
unexpected crimes, we usually install various monitors. The existence of
various monitors plays a role of continuous observation, hence awe potential
violators away from deregulation. From this point of view, our work clarified
such scenarios in the classical world in terms of the famous paradox proposed
by Zeno.

We thank helpful discussions with L. G. Wang, W. Q. Chen, and H. Q. Lin. This
work is supported by the Earmarked Grant Research from the Research Grants
Council of HKSAR, China (Project No. CUHK 400807).


\begin{thebibliography}{9}
\bibitem{Mario} Super Mario is the protagonist in the popular game of Super
Mario.
\end{thebibliography}
\end{document}